\documentclass[reprint]{revtex4-1}

\usepackage{colortbl}
\usepackage{float}
\usepackage{dcolumn}% Align table columns on decimal point
\usepackage{color}
\usepackage{mathtools}
\usepackage{subfig}
\usepackage{tabularx} % extra features for tabular environment
\usepackage[final]{hyperref} % adds hyper links inside the generated pdf file
\usepackage{blindtext}
\usepackage{xcolor}
\usepackage[normalem]{ulem}               % to striketrhourhg text
\newcommand\blueout{\bgroup\markoverwith
{\textcolor{blue}{\rule[0.5ex]{2pt}{0.8pt}}}\ULon}
\usepackage{xr}
\usepackage{soul}
\usepackage[version=4]{mhchem}
\usepackage{siunitx}

\begin{document}

\title{Relativistic Embedded Equation-of-Motion Coupled-Cluster Approach to the Core-Ionized States of Actinides: A Case Study of Uranyl(VI) in \ce{Cs2UO2Cl4}}

\author{Wilken Aldair Misael}
\author{Andr\'{e} Severo Pereira Gomes}
\altaffiliation{Author to whom correspondence should be addressed}  
\email{andre.gomes@univ-lille.fr}
\affiliation{Univ. Lille, CNRS, UMR 8523-PhLAM-Physique des Lasers Atomes et Mol\'{e}cules, F-59000 Lille, France}

\date{\today}

\begin{abstract}

We investigate the core-level ionization energies of the bare uranyl ion (\ce{UO2^{2+}}) and its interaction with X-rays when it is hosted in the \ce{Cs2UO2Cl4} crystalline environment using a recent implementation of the core-valence-separated relativistic equation-of-motion coupled-cluster method (CVS-EOM-CC). Our study evaluates different relativistic Hamiltonians, % and spinors, 
assesses basis set and virtual space truncation effects, and examines the role of orbital correlation and relaxation in simulating the spectroscopic observables. The results of this investigation highlight the importance of computing two-electron interactions beyond the zeroth-order truncation (i.e. the Coulomb term) when working in the tender and hard X-ray ranges. Additionally, we compare different structural models using the frozen density embedding method (FDE). By contrasting the bare and embedded uranyl models, we observe significant changes in binding energies, highlighting the influence of the equatorial ligands of the uranyl ion on its spectroscopic observables. A comparison between the embedded uranyl and supermolecular systems, excluding the cesium atoms, reveals systematic differences, with binding energy variations from experimental data remaining within \qty{10}{\electronvolt}. Notably, the computed spin-orbit splittings for U $4d$ and $4f$ deviate by less than \qty{0.7}{\electronvolt}, demonstrating the validity of this protocol for computing binding energies in the soft X-ray range.

\end{abstract}

\maketitle

\section{Introduction}
\label{sec:intro}

Photoemission spectroscopies, including ultraviolet photoelectron spectroscopy (UPS), developed by Turner and collaborators \cite{turner1970molecular}, X-ray photoelectron spectroscopy (XPS), pioneered by Siegbahn \cite{siegbahn1982electron}, and angle-resolved photoemission spectroscopy (ARPES) \cite{borisenko202199,zhang2022angle}, are well-established \textit{photon-in electron-out} techniques rooted in the photoelectric effect \cite{hertz1887ueber, einstein1905indeed}. These techniques enable the investigation of various physical and chemical properties, such as the electronic structure of materials, through the ionization of electrons by monochromatic light. 

The application of photoemission spectroscopy to actinide-containing materials is particularly insightful, offering valuable information on their chemical bonding, coordination chemistry, and oxidation states of the 5\textit{f} elements \cite{kotani1993theory,bagus2013theory,drot1998structural,dau2012photoelectron,baker2014uranium,perry2015tris,rout2017spectroscopic,sun2018uranium,maslakov2019nature,el2022monitoring,thompson2024review}, with valence and core spectra providing complementary information\cite{kvashnina2013chemical,butorin2022x,fujimori2019manifestation,teterin2016valence}. However, interpreting the photoemission spectra in general (and that of of actinide materials in particular) presents challenges due to various factors, including significant electronic correlation, ligand field effects on multiplet structure, satellite peaks, and vibronic coupling observed in the recorded spectra \cite{kotani1993theory, ilton2008ligand, bagus2013theory, ilton2011xps, li2014strong, su2015photoelectron, li2017probing, gibson2018experimental}. In valence-level experiments, spectral interpretation is further complicated by substantial peak overlap due to the high density of states. This makes it difficult to distinguish the partial cross-sections of different excitations and decay processes, as they energetically coincide~\cite{kotani1993theory, bagus2013theory, teterin2016valence}. In this regard, XPS benefits from the element-specific and orbital-selective characteristics of X-ray spectroscopies, as the energy levels of core orbitals fall precisely within the X-ray range \cite{norman2018simulating}, thereby simplifying the complexity of the spectra to be interpreted.

Given the complexity of the physical processes involved, theoretical modeling at the atomistic level has become indispensable to gain a deeper understanding of the interactions between actinides and X-ray and support the interpretation of experimental data~\cite{ilton2008ligand, ilton2011xps, bagus2021computation, bagus2022origin}. The accurate treatment of core levels, as well as valence levels for heavy elements such as actinides, requires three key ingredients: first, the description of relativistic effects on the electronic structure~\cite{pyykko1979relativity,pyykko1988relativistic,pyykko2004theoretical,pyykko2012relativistic,pyykko2012physics}. These are most reliably taken into account by solving the Dirac equation\cite{liu2012perspectives,liu2014perspective,liu2020essentials,liu2021relativistic}, and give rise to spin-orbit coupling (SOC) as well as the contraction (expansion) of \textit{s} and \textit{p} (\textit{d} and \textit{f}) orbitals due to scalar relativistic (SR) effects. 

Second, a treatment of electron correlation--and for core states the relaxation effects arising from the creation of core holes--which are very important for accurately describing the energetics of excitation or ionization processes~\cite{infante2006performance,schreckenbach2010theoretical,tecmer2012charge,tecmer2011electronic,gomes2013towards,gomes2015applied,boguslawski2017multi,gibson2018experimental,oher2020influence,oher2020investigation,oher2021influence,liu2021relativistic}. For actinides and in particular the uranyl (VI) species\ce{UO2^{2+}}, which is ubiquitous in the chemistry of actinides in the environment (solution, etc), valence processes have been extensively studied over the last two decades, with different electron correlation approaches such as coupled cluster (CC)\cite{infante2006performance,gomes2013towards, tecmer2014communication,gomes2015applied,real2009benchmarking}, complete active space second order perturbation theory (CASPT2) \cite{real2007theoretical, pierloot2005electronic,pierloot2005electronic,van2006electronic, boguslawski2017multi} and density functional theory (DFT)~\cite{vitova2017role,oher2023does}. One key finding from these prior investigations is that that \ce{UO2^{2+}} and its complexes with ligands such as chlorides, nitrates, carbonates etc are well-represented by a single reference in their electronic ground state.  

More recently there has been growing interest in accurate simulations of core states, notably employing multireference methods like restricted active-space self-consistent field~(RASSCF)~\cite{sergentu2018ab,sergentu2022x,polly2021relativistic,stanistreet2023bounding,stanistreet2024quantifying} and multireference configuration interaction (MRCI)~\cite{ehrman2024unveiling} as well as DFT~\cite{konecny2021accurate,misael2023core}. For ionizations, we note the use of the $\Delta$SCF or $\Delta$MP2 approaches\cite{south20164}, which can be very effective in recovering electron correlation and relaxation effects in the calculation of core electron binding energies (CBEs), but at the cost of performing SCF calculation for different electronic configurations, which may not be trivial to converge in certain cases. In view of the closed shell character of uranyl (VI) systems, an alternative to these $\Delta$ methods for ionizations is the equation of motion coupled cluster method for ionization energies (EOM-IP-CC) within the core-valence separation (CVS) approximation\cite{coriani2015communication}, which consists of removing unwanted contributions from valence as well as high-lying unbound states to the EOM similarity transformed Hamiltonian (and trial vectors)\cite{vidal2019new,halbert2021relativistic} though projection. Within this framework, on top of accounting for electron correlation orbital relaxation due to the creation of the core hole(s) is incorporated through the excitation operator $\hat{R}$ \cite{bartlett2012coupled, helgaker2012recent,halbert2023}. CVS-EOM-CC approaches have shown very good accuracy for systems containing light elements\cite{coriani2015communication,vidal2019new,vidal2020equation,fransson2013carbon,schnack2023new}--often showing differences of less than \qty{1}{\electronvolt} in comparison to experiment--as well as for heavier elements such as iodine or xenon~\cite{southworth2019observing,halbert2021relativistic}, but have not been explored to date, and to the best of our knowledge, for actinide systems.

The third and final ingredient required is a description of the interaction between the center of interest and its environment. This can be done by considering a sufficiently large (``supermolecular'') system and treating it at a given level of theory. This route poses difficulties to the application of wavefunction-based correlated electronic structure methods due to their steep scaling with system size, which are not as easily alleviated by the development of reduced scaling approaches~\cite{crawford2019reduced} for the case of excitations or ionization as for the description of the ground state, or by the use of massively parallel implementations~\cite{pototschnig2021implementation}.

A better alternative is to employ embedding methods, in which the total systems is subdivided into (interacting) subsystems, with one or a few subsystems of interest being treated with computationally expensive approaches such as coupled cluster, whereas the remaining ones are treated at a less sophisticated level (e.g.\ mean-field theories such as Hartree-Fock or DFT, or with classical descriptions), thereby reducing the overall computational cost. Among the different embedding methods employed to heavy elements in general and actinides in particular, we note the frozen density embedding (FDE) method\cite{wesolowski1993frozen,wesolowski2015frozen}, which is appealing as it provides a quantum mechanical description for all subsystems, and represents their interaction through a local embedding potential. To date, FDE has been combined time-dependent DFT (TDDFT) to obtain X-ray Absorption Near Edge Structure (XANES) spectra~\cite{misael2023core} of uranyl species in the \ce{Cs2UO2Cl4} crystal (a prototypical host system for actinyls, for which there is a wealth of experimental data \cite{denning1976electronic,denning2002covalency,denning2007electronic,vitova2015polarization,teterin2016valence,stanistreet2024quantifying}). However, FDE remains unexplored to investigate CBEs with CVS-EOM-CC theory, though we note such combination has been employed to obtain CBEs of chloride species in water droplets and adsorbed on ice surfaces~\cite{opoku2022simulating}. 

The goal of this work is therefore to explore combination of FDE and CVS-EOM-CC to obtain CBEs for different levels of uranium ($1s, 2s, 2p, 3s, 3p, 3d, 4s, 4p, 4d, 4f$) and the $1s$ level of oxygen atoms in the \ce{Cs2UO2Cl4} crystal. Through this study we intend to provide, first, information on the most suitable computational setup (choice of Hamiltonian, truncation of the virtual space etc) for different core leve uranyl, following up on the preliminary investigations of relativistic CVS-EOM-CC on heavy elements by~\citet{halbert2021relativistic}, which have indicated the ability of the method to achieve high accuracy for absolute CBEs.  Second, we aim to compare our theoretical results to the photoemission measurements conducted by \citet{teterin2016valence} for the \ce{Cs2UO2Cl4} crystal which, to the best of our knowledge, reports the sole experimental investigation of CBEs for uranyl. We note that to the best of our knowledge, CBEs for the uranyl $2p$ levels had been investigated by \citet{south20164} at the $\Delta$SCF and $\Delta$MP2 levels, but only for the bare uranyl ion which is not experimentally accessible, and for which calculations were made computationally less demanding by exploiting double point group symmetry to a degree which is not generally feasible for realistic systems.

The manuscript is organized as follows: in Section \ref{sec:compdet} we present the computational details of our calculations. In Sections \ref{sec:rel_cont} and  \ref{sec:dyn-cor}, we investigate the impact of different Hamiltonians and correlating spaces on the CBEs of the bare uranyl ion in order to arrive at the best balance between cost and accuracy in the relativistic CVS-EOM-CC treatment for all CBEs of interest. In Section \ref{sec:env-eff}, we proceed to exploring the effect of equatorial ligands on the CBEs, following the prescription by~\citet{gomes2013towards} in which we compare the performance of structural models in which the chloride ligands situated at the equatorial plane of uranyl are either represented via an embedding potential or explicitly included in the correlated calculations, and compare our results to the experimental work by~\citet{teterin2016valence}, followed by our conclusions and perspectives for further work.

\section{Computational Details}
\label{sec:compdet}

CBEs have been calculated using relativistic-based equation-of-motion coupled-cluster calculations at the singles and doubles level (EOM-IP-CCSD)\cite{shee2018equation}, employing the core-valence-separation (CVS) approximation\cite{halbert2021relativistic} (CVS-EOM-IP-CCSD) to compute the ionization channels of interest (O $1s$ and U $1s, 2s, 2p, 3s, 3p, 3d, 4s, 4p, 4d, 4f$). Furthermore, we utilized the frozen density embedding (FDE) method \cite{wesolowski1993frozen,wesolowski2015frozen} to investigate the role of environmental effects in the computation of these spectroscopic observables, which we will refer to as CVS-EOM-IP-CCSD-in-DFT.

CVS-EOM-IP-CCSD calculations and CVS-EOM-IP-CCSD-in-DFT calculations were performed in DIRAC22~\cite{DIRAC22} version of the \texttt{DIRAC} electronic structure code~\cite{saue2020dirac}, as well as development snapshots (\texttt{34fbd49,4b35e48,d70bbe283,e061718,e0617189f6, \\ e7e2094}). We employed Dyall's all-electron basis sets of double and triple-zeta quality \cite{dyall2002relativistic,dyall2016relativistic} for uranium and Dunning's cc-pVTZ basis set \cite{denning1976electronic} for all other atoms. These basis sets were left uncontracted. We have employed a Gaussian nuclear model in all calculations \cite{visscher1997dirac}.

For EOM-CC calculations, we employed the Dirac-Coulomb Hamiltonian ($^4$DC), the Exact 2-Component molecular mean-field (X2Cmmf)~\cite{sikkema2009molecular} Hamiltonian ($^{2}$DC$^{M}$), in which the decoupling of the large and small components of 4-spinors is achieved by block diagonalizing the converged 4-component Fock matrix at the end of the SCF procedure. In benchmark calculations with EOM-IP-CCSD, we have also employed the X2Cmmf-Gaunt~
\cite{sikkema2009molecular} Hamiltonian ($^{2}$DCG$^{M}$), in which the Gaunt integrals are included in the construction of the (4-component) Fock matrix of the SCF step, but are not explicitly included in the transformation from atomic to molecular basis. Apart from the case of CVS-EOM-IP calculations on the bare uranyl ion, which was used to investigate all aforementioned channels of interest, for less energetic channels in which the (SS$\mid$SS)-type integrals did not strongly contribute to the CBEs, these were by a simple Coulombic correction~\cite{visscher1997approximate} for reasons of computational cost. Also, as will be further discussed in more detail, for CVS-EOM-IP calculations we have explored various correlation spaces. 

\begin{figure*}
    \centering
    \includegraphics[width=0.65\linewidth]{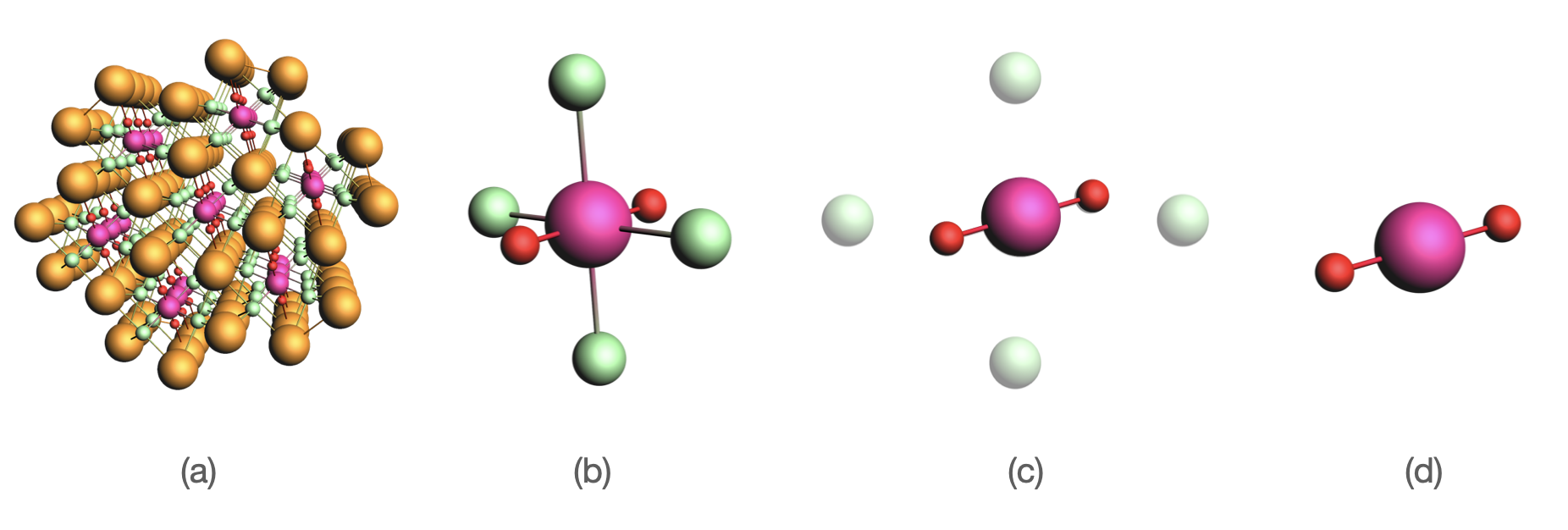}
    \caption{Reference system for this study: (a) Dicesium uranyl(VI) tetrachloride crystal (\ce{Cs2UO2Cl4}). Models here investigated: (b) uranyl(VI) tetrachloride dianion, \ce{UO2Cl4^{2-}} (c) uranyl(VI) ion in the FDE embedding potential of four chloride atoms, \ce{UO2^{2+}} @ \ce{Cl4^{4-}} and (d) bare uranyl(VI) ion, \ce{UO2^{2+}} (cesium: orange; uranium: pink; oxygen: red; chlorine: green).}
    \label{fig:models}
\end{figure*}

In all of our calculations, the structures employed were based on the experimental crystal structure of \ce{Cs2UO2Cl4} reported by~~\citet{watkin1991structure}, whose U-O and U-Cl bond lengths are 1.774~\AA \ and 2.673~\AA \ respectively. We employed the same structural models and subsystem partitioning for the embedding calculations as outlined by~~\citet{gomes2013towards}. All calculations on the bare uranyl and uranyl tetrachloride were carried out in D$_{\infty_h}$ and $D_{2h}$ symmetry, respectively, while for embedded uranyl simulations the crystalline site symmetry (C$_{2h}$) was used. These systems are represented in Figure~\ref{fig:models}. The embedding potentials employed here were obtained from freeze-thaw calculations employing the scalar-relativistic ZORA Hamiltonian \cite{van1996zero}, TZ2P basis sets, the PW91k kinetic energy \cite{burke1998derivation}, the PBE exchange-correlation functional \cite{ernzerhof1999assessment} for the non-additive terms and subsystem energies. These calculations were carried out via the \texttt{PyADF} scripting framework \cite{jacob2011pyadf} and the embedding potentials were subsequently imported into the \texttt{DIRAC} calculations. 

In the comparison to experiment, we accounted for the Breit and leading QED effects (self interaction and vacuum polarization) on the uranium orbital energies through atom-based energy corrections to the CVS-EOM-IP-CCSD $^{2}$DC$^{M}$ calculations, based on the work of~~\citet{koziol2018qed}, as done elsewhere in the literature~\cite{southworth2019observing,halbert2021relativistic}. Since these are not available for uranium, we obtained values by interpolation. 

The data corresponding to the results in this paper is available at the Zenodo repository~\cite{dataset-xas-uranyl}.

\section{Results}

In this section, we present our results on the oxygen $1s$ and various uranium CBEs at different edges. Building on our previous work on uranyl ion core-level absorption spectra, we aimed to evaluate the FDE method's ability to describe the influence of equatorial ligands on its core-level spectroscopic observables. To this end, we used the XPS measurements of the \ce{Cs2UO2Cl4} crystal from \citeauthor{teterin2016valence}~\cite{teterin2016valence} as a reference in our discussion. Given the limited experimental data available for uranyl tetrachloride (restricted to O $1s$ and U $4p$, $4d$, and $4f$ binding energies), we additionally aimed to investigate, at a fundamental level, the extent of relativistic effects on these spectroscopic observables of uranyl across different energy ranges. Specifically, we have also calculated the U $1s$, $2s$, $2p$, $3s$, $3p$, and $3d$ binding energies for the bare uranyl ion. For each binding energy, we evaluated the performance of different Hamiltonians and spinors used to compute these quantities. Additionally, we examined the effects of orbital relaxation and correlation on the results, as described below. % Due to constraints in computational resources, we only employed double-zeta basis sets in calculations for the dianion. Bare and embedded uranyl calculations were performed using double- and triple-zeta basis sets.

\subsection{Assessment of relativistic Hamiltonians}
\label{sec:rel_cont}

\subsubsection{The performance of two-component approaches}

Table \ref{table*:cvs-eom-ip_benchmark} presents the results of our investigation, including the findings from 4-component calculations (denoted as $^4$DC) and the variations observed when employing other Hamiltonians. Our results indicate that the discrepancies in binding energies between the reference $^4$DC calculations and the computationally more efficient $^2$DC$^M$ calculations are generally negligible, with values of approximately 0.3 eV or less in absolute terms. However, for the uranium $1s$ binding energy, the differences reach approximately 2.4 eV in absolute value. The observed discrepancy in our results follows a trend observed by \citeauthor{halbert2021relativistic}~\cite{halbert2021relativistic} for halides, where it increases modestly with atomic number, being approximately 1.7 eV for astatide. 

%Our findings align with the results reported by \citeauthor{repisky2023modern}, especially concerning the observed behavior for Cn \cite{repisky2023modern}. In their study, they also reported the application of advanced X2C approaches, which exhibited a notable reduction in the PCE for the calculated properties.

\begin{table*}%[!htbp]
  \caption{CVS-EOM-IP ($^4$DC) calculation results and $\Delta E$ differences (in eV) between the reference $^4$DC values and various relativistic Hamiltonians for uranium and oxygen in \ce{UO2^{2+}}. Hamiltonians are discussed in the text. Uncontracted Dyall double-zeta basis sets~\cite{dyall2016relativistic,dyall2002relativistic} are used for all atoms. It should be noted that different values for a given edge correspond to splitting caused by the reduction in symmetry in the molecular system compared to the isolated atoms.}
  \label{table*:cvs-eom-ip_benchmark}
 % \resizebox{\textwidth}{!}{%
  \begin{tabular}{@{}llcccccc@{}}
  \hline\hline
   \multicolumn{2}{c}{Assignment}    &   $^4$DC         &     \multicolumn{5}{c}{Difference with respect to the $^4$DC reference}                  \\
  \cline{4-8}
  &  &  & $^4$DC & $^2$DC$^M$  & $^2$DC$^M$ & $^2$DCG$^M$   & $^2$DCG$^M$  \\
  &  & & (SS$\mid$SS) & & (SS$\mid$SS) & & (SS$\mid$SS) \\
  \hline
  U    & $1s$  & 116432.54          & -68.44               & -2.42 & -69.57                & -525.14 & -591.68                 \\
       & $2s$  & 21905.11           & -12.56               & -0.14 & -12.58                & -64.17  & -76.72                  \\
       & $2p_{1/2}$  & 21095.75           & -17.97               & 0.02  & -17.83                & -107.87 & -125.57                 \\
       & $2p_{3/2}$  & 17257.06           & -5.86                & -0.28 & -6.08                 & -68.52  & -74.27                  \\
       &        & 17256.97           & -5.85                & -0.28 & -6.07                 & -68.51  & -74.26                  \\
     & $3s$          & 5609.00            & -3.16                & $< 3 \times 10^{-3}$  & -3.14               & -12.68  & -15.80                  \\
  & $3p_{1/2}$      & 5239.69            & -4.10                & 0.02  & -4.08                 & -21.44  & -25.51                  \\
  & $3p_{3/2}$      & 4348.50            & -1.52                & $< 3 \times 10^{-2}$ & -1.54                 & -13.13  & -14.64                  \\
       &            & 4347.97            & -1.53                & -0.03 & -1.55                 & -13.13  & -14.64                  \\
       & $3d_{3/2}$      & 3765.71            & -1.43                & 0.02  & -1.40                 & -9.52   & -10.92                  \\
       &            & 3765.45            & -1.42                & 0.02  & -1.39                 & -9.50   & -10.90                  \\
       & $3d_{5/2}$      & 3585.16            & -0.21                & -0.05 & -0.26                 & -6.71   & -6.92                   \\
       &            & 3584.86            & -0.22                & -0.06 & -0.27                 & -6.70   & -6.92                   \\
       &            & 3584.72            & -0.22                & -0.06 & -0.27                 & -6.70   & -6.91                   \\
       & $4s$        & 1480.55            & -0.83                & 0.01  & -0.82                 & -2.93   & -3.76                   \\
       & $4p_{1/2}$      & 1313.23            & -1.03                & 0.01  & -1.03                 & -5.03   & -6.06                   \\
       & $4p_{3/2}$      & 1079.22            & -0.37                & $< 2 \times 10^{-3}$  & -0.37                 & -2.79   & -3.16                   \\
       &            & 1078.20            & -0.37                & $< 1 \times 10^{-4}$  & -0.37                 & -2.78   & -3.15                   \\
       & $4d_{3/2}$      & 812.39             & -0.30                & $< 2 \times 10^{-3}$  & -0.29                 & -1.59   & -1.89                   \\
       &            & 811.47             & -0.29                & 0.01  & -0.29                 & -1.58   & -1.87                   \\
       & $4d_{5/2}$      & 768.84             & -0.02                & -0.01 & -0.02                 & -0.92   & -0.93                   \\
       &            & 768.06             & -0.01                & $< 2 \times 10^{-3}$   & -0.01                 & -0.91   & -0.92                   \\
       &            & 767.59             & -0.01                & $< 3 \times 10^{-3}$   & -0.01                 & -0.90   & -0.92                   \\
       & $4f_{5/2}$      & 417.08             & 0.05                 & 0.01  & 0.05                  & 0.06    & 0.10                    \\
       &            & 416.36             & 0.04                 & $< 6 \times 10^{-4}$   & 0.05                  & 0.06    & 0.10                    \\
       &            & 416.01             & 0.05                 & 0.01  & 0.05                  & 0.07    & 0.12                    \\
       & $4f_{7/2}$      & 406.02             & 0.06                 & $< 5 \times 10^{-4}$   & 0.06                  & 0.08    & 0.13                    \\
       &            & 405.39             & 0.05                 & 0.01  & 0.05                  & 0.07    & 0.12                    \\
       &            & 404.90             & 0.05                 & 0.01  & 0.05                  & 0.07    & 0.12                    \\
       &            & 404.70             & 0.05                 & 0.01  & 0.05                  & 0.07    & 0.12                    \\
     \hline
  O     & $1s$         & 555.52             & 0.01                 & 0.01  & 0.01                  & 0.03    & 0.05                    \\
       &            & 555.52             & 0.01                 & 0.01  & 0.01                  & 0.03    & 0.05                    \\
  \hline\hline
  \end{tabular}%}
  \end{table*}

Our results further illustrate an observation on the magnitude the two-electron picture change error (here denoted as PCE, and shown by the difference between a reference 4-component calculation and a 2-component one) for CBEs, first analyzed by~\citet{halbert2021relativistic} for the molecular mean-field ($^2$DC$^M$) approach and subsequently by \citet{repisky2023modern} for atomic mean-field approaches, . If PCE does grow in absolute value as one goes down in the periodic table, and can reach a few electron volts for the deepest channels ($1s$)--here, PCE is 2.42 eV in absolute value for U $1s$, though it is not larger than 0.3 eV in absolute value for all other channels--we see that in relative terms, this error represents only 0.0021\% of the calculated CBE for  U $1s$ and even in absolute value this error is dwarfed by the magnitude of other contributions to the Hamiltonian such as those we discuss in the following. That being said, such errors must be taken into consideration in a comparison to experiment, should these ever be available for such deep cores.

However, it is important to note that this conclusion may not be generalized to other properties. \citet{repisky2023modern} acknowledges that although a significant portion of the PCE error can be recovered by employing their X2C approaches, they still observed qualitative discrepancies. For instance, they highlight their results for the contact density contribution of the Cn $1s$ shell in CnF$_6$, which experiences a shift of approximately 11\% between the 2- and 4-component frameworks. And to the best of our knowledge, no attempt to date has been made to investigate their importance to intensities.

\subsubsection{The role of (SS\textbar{}SS)-type  integrals}

We also examined the significance of explicitly calculating two-electron integrals of (SS\textbar{}SS)-type in the simulation simulating core-level spectroscopic properties of actinides, building upon the previous work of \citeauthor{south20164} on uranium $2p$ binding energies \cite{south20164}. The computation of these integrals is computationally demanding, and typically, the Coulombic correction method proposed by  \citeauthor{visscher1997approximate}~\cite{visscher1997approximate} is employed to alleviate the computational demands.

Based on our findings, also displayed in Table \ref{table*:cvs-eom-ip_benchmark}, it is evident that the inclusion of these integrals has a significant role when computing the innermost ionization energies of uranium. In addition, we also observe that the explicit inclusion of (SS\textbar{}SS)-type integrals has a more pronounced effect on $p_{1/2}$ spinors compared to other spinors within the same shell. These findings align with the observations made by \citet{halbert2021relativistic} for astatide.

Furthermore, our results indicate that these integrals generally increase the ionization energies above the $4f_{5/2}$ level, while for lower levels, the opposite trend is observed, with only modest variations. Finally, our calculated uranium $2p_{1/2}$ and $2p_{3/2}$ ionization energies, including (SS\textbar{}SS)-type integrals, show quantitative agreement with the results reported by \citet{south20164}, despite the differences in theory level ($\Delta$HF/dyall.v3z). 

\subsubsection{Electron-electron interaction beyond the Coulomb term}

Additionally, we explored the influence of current-current interactions on these spectroscopic observables, also shown in Table \ref{table*:cvs-eom-ip_benchmark}. Specifically, we included the Gaunt interaction in the two-electron operator, thereby taking into account the contributions from spin-other-orbit (SOO) interactions.

From our results, we observe that these interactions exhibit significantly larger magnitudes than those from (SS\textbar{}SS)-type integrals. For instance, the uranium 1s ionization energy shows a contribution of 522.7 eV from the Gaunt interaction, whereas the (SS\textbar{}SS)-type integrals contribute only 67.1 eV. Similarly, for the uranium $2s$ ionization energy, the Gaunt interaction contributes 68 eV, while the (SS\textbar{}SS)-type integrals contribute 6 eV. 

These results are consistent with the findings of \citet{south20164} regarding uranium $2p$ binding energies, as well as the results of \citet{halbert2021relativistic} and \citet{repisky2023modern} on the inner-core edges of  other heavy elements. They underscore the relevance of considering the electron-electron repulsion beyond the zeroth order truncation, i.e., the Coulomb interaction.

\subsection{Dynamical correlation and orbital relaxation effects}
\label{sec:dyn-cor}

Table \ref{table*:koop-cvs} displays the contributions of electron correlation and orbital relaxation due to the formation of the core hole for CBEs of \ce{UO2^{2+}} calculated using the CVS-EOM-IP method. These contributions were determined by subtracting the CVS-EOM-IP values from those obtained using Koopmans' theorem. Significant discrepancies are observed between the Koopmans' theorem and CVS-EOM-IP calculations, regardless of the Hamiltonian used. As expected, the influence of dynamical correlation and orbital relaxation becomes increasingly prominent for inner shells, ranging from 19 eV for the O $1s$ binding energy to over 100 eV for the U $1s$ binding energy. These findings are consistent with the investigation conducted by \citeauthor{south20164} on $2p$ spinors using the $\Delta$HF and $\Delta$MP2 methods, where the former accounts for orbital relaxation and the latter also recovers orbital correlation.

The results demonstrate that the CVS-EOM-IP framework offers an alternative for exploring the effects of orbital correlation and relaxation in systems containing elements beyond the second row of the periodic table. Furthermore, this approach should provide a more reliable alternative for obtaining such excited states compared to methods that optimize the wave function of the ground and core-excited states, such as the $\Delta$MP2 method. It has been shown by \citeauthor{ljubic2014reliability}~\cite{ljubic2014reliability} and \citeauthor{arias2022accurate}~\cite{arias2022accurate} that the $\Delta$MP2 method often overestimates correlation energy and increases ionization energies, with additional convergence issues and an overall computational effort. Given that these limitations were observed for systems containing lighter elements, it is reasonable to anticipate the same behavior in the computation of excited states for heavier elements as well.

\begin{table*}%[!t]
	\caption{Differences between the CVS-EOM-IP CBEs and the corresponding Koopmans' values (in eV) are presented for \ce{UO2^{2+}} using the uncontracted Dyall double-zeta basis set for all atoms.}
	\label{table*:koop-cvs}
	%\resizebox{\textwidth}{!}{%
    \begin{tabular}{llcccccc} 
    \hline\hline
      \multicolumn{2}{c}{Assignment} & $^4$DC    & $^4$DC & $^2$DC$^M$   & $^2$DC$^M$ & $^2$DCG$^M$  & $^2$DCG$^M$  \\ 
      &  &     & (SS\textbar{}SS)  &    & (SS\textbar{}SS) &   & (SS\textbar{}SS)  \\ 
    \hline
  U    & $1s$          & 102.96 & 104.18 & 105.37 & 105.31 & 105.09 & 104.96  \\
         & $2s$         & 66.68  & 66.72  & 66.81  & 66.74  & 66.90  & 67.20   \\
         & $2p_{1/2}$         & 70.31  & 70.32  & 70.28  & 70.18  & 70.15  & 70.16   \\
         & $2p_{3/2}$         & 65.58  & 65.53  & 65.67  & 65.75  & 65.61  & 65.64   \\
         & $3s$         & 38.98  & 38.87  & 38.98  & 38.86  & 38.87  & 39.00   \\
         & $3p_{1/2}$         & 41.48  & 41.50  & 41.46  & 41.48  & 41.43  & 41.42   \\
         & $3p_{3/2}$         & 38.37  & 38.39  & 38.39  & 38.41  & 38.44  & 38.45   \\
         & $3d_{3/2}$         & 41.15  & 41.21  & 41.13  & 41.18  & 41.13  & 41.18   \\
         & $3d_{5/2}$         & 41.09  & 41.03  & 41.14  & 41.08  & 41.08  & 41.20   \\
         & $4s$         & 20.97  & 20.98  & 20.96  & 20.98  & 20.90  & 20.92   \\
         & $4p_{1/2}$         & 21.75  & 21.70  & 21.75  & 21.69  & 21.62  & 21.56   \\
         & $4p_{3/2}$         & 20.08  & 20.04  & 20.08  & 20.05  & 20.01  & 19.97   \\
         & $4d_{3/2}$         & 19.92  & 19.95  & 19.92  & 19.94  & 19.88  & 19.90   \\
         & $4d_{5/2}$         & 19.70  & 19.71  & 19.70  & 19.71  & 19.79  & 19.80   \\
         & $4f_{5/2}$         & 19.53  & 19.58  & 19.53  & 19.57  & 19.56  & 19.51   \\
         & $4f_{7/2}$         & 19.45  & 19.35  & 19.45  & 19.36  & 19.42  & 19.47   \\ 
    \hline
    O     & $1s$          & 19.19  & 19.18  & 19.19  & 19.18  & 19.03  & 19.02   \\
    \hline\hline
    \end{tabular}%}
    \end{table*}

\subsubsection{The effect of virtual space truncation}

To address the high computational demands of these simulations when working with larger systems, as we will in the following section, we investigated the impact of using various cutoffs for the correlated virtual spinor space. Table~\ref{table*:vir-spc} presents the results of our investigation, where we utilized the less expensive $^2$DC$^M$ Hamiltonian and a double zeta basis set for the bare uranyl ion in its pristine D$_{\infty h}$ symmetry. These findings reveal significant variations, approximately 39 eV, in the binding energy of the uranium $1s$ orbital when transitioning from the full virtual space to a truncation that retains approximately 60\% of the virtual space. 

%This result is in agreement with the study by  \citeauthor{halbert2021relativistic}~\cite{halbert2021relativistic}, who observed that correlating high-lying virtual orbitals is necessary to obtain quantitative results when studying inner shells of heavy elements.

Additionally, noticeable deviations, reaching up to \qty{10}{\electronvolt} for the uranium $2p_{3/2}$ binding energy, are observed. However, for lower energy levels, such truncations of the virtual space do not introduce significant errors, with deviations of approximately 1 eV or less when compared to calculations involving the full virtual space. Consequently, we have determined that it is feasible to accurately investigate uranium binding energies from the $4f_{3/2}$ level up to the $3s$ level by limiting our correlation space to spinors with energies ranging from -200 to 200 E$_h$. This approach encompasses approximately 61.4\% of the available virtual orbitals when employing a double zeta basis set for the bare uranyl ion, and was the one applied in the subsequent investigation.

\begin{table*}[!ht]
    \caption{Results from CVS-EOM-IP calculations for CBEs (in eV) in \ce{UO2^{2+}} in which all orbitals are included in the correlation space, and the differences with this reference when employing different correlation spaces. All calculations were performed with the $^2$DC$^M$ Hamiltonian and using an uncontracted Dyall double-zeta basis set for all atoms.}
    \label{table*:vir-spc}
    \small % Use a smaller font size for the table
    \begin{tabular*}{0.8\linewidth}{@{\extracolsep{\fill}}llcccccc@{}}
        \hline\hline
        \multicolumn{2}{c}{Assignment} & $^2$DC$^M$ & \multicolumn{5}{c}{Percentage of correlated virtual orbitals} \\
        \cline{4-8}
         & & & 72.5 & 67.6 & 62.5 & 61.4 & 56.7 \\
        \hline
        U & $1s$ & 116430.12 & 5.19 & 11.91 & 25.64 & 25.76 & 38.5 \\
        & $2s$ & 21904.97 & 0 & -0.35 & 0.8 & 4.47 & 4.61 \\
        & $2p_{1/2}$ & 21095.77 & 0 & -0.71 & 0.53 & 5.56 & 5.78 \\
        & $2p_{3/2}$ & 17256.74 & -0.58 & 0.22 & 4.24 & 4.42 & 10.62 \\
        & $3s$ & 5609.00 & -0.18 & -0.14 & 0 & 0 & 0.98 \\
        & $3p_{1/2}$ & 5239.71 & -0.26 & -0.33 & -0.05 & 0.07 & 1.42 \\
        & $3p_{3/2}$ & 4348.21 & -0.21 & -0.27 & -0.08 & -0.01 & 0.87 \\
        & $3d_{3/2}$ & 3765.60 & -0.06 & -0.34 & -0.26 & -0.38 & 0.9 \\
        & $3d_{5/2}$ & 3584.86 & -0.05 & -0.3 & -0.41 & -0.35 & 0.83 \\
        & $4s$ & 1480.56 & -0.06 & -0.05 & -0.06 & -0.06 & -0.08 \\
        & $4p_{1/2}$ & 1313.24 & -0.08 & -0.12 & -0.14 & -0.15 & -0.05 \\
        & $4p_{3/2}$ & 1078.71 & -0.06 & -0.09 & -0.12 & -0.11 & -0.12 \\
        & $4d_{3/2}$ & 811.93 & -0.02 & -0.11 & -0.16 & -0.24 & -0.26 \\
        & $4d_{5/2}$ & 768.16 & -0.01 & -0.1 & -0.23 & -0.23 & -0.26 \\
        & $4f_{5/2}$ & 416.49 & 0.01 & 0.02 & 0.05 & 0.05 & 0.04 \\
        & $4f_{7/2}$ & 405.25 & 0.01 & 0.02 & 0.03 & 0.05 & 0.04 \\
        \hline
        O & 1s & 555.52 & 0 & 0 & -0.03 & -0.03 & -0.26 \\
        \hline\hline
    \end{tabular*}
\end{table*}

  \subsection{Environment effects on uranyl CBEs}
  \label{sec:env-eff}

\subsubsection{The effect of approximately representing the equatorial ligands}  
  % As previously discussed, due to the limitations of employing uncontracted basis sets and the constraints on our computational resources, we could only perform calculations with double-zeta bases for \ce{UO2Cl4^{2-}}. Following the discussion in the previous section, we restricted the virtual space to spinors with energies up to approximately 200 E$_h$. To ensure an unbiased comparison, we applied a similar constraint to the calculations for the bare and embedded uranyl models. 
  
Table \ref{table*:cvs-eom-ipmodels} presents the ionization energy values obtained from CVS-EOM-IP calculations for the three models shown in Figure \ref{fig:models}, studied with both double- and triple-zeta basis sets. At the double-zeta level, it is observed that the transition from the bare to the embedded uranyl results in significant changes in the binding energies, with a decrease of nearly 21.5 eV when moving from the bare to the embedded case. 
  
  When improving the basis set while keeping the correlation space unchanged, greater deviations are observed in the case of the bare uranyl system, reaching up to 1.4 eV, whereas the embedded case exhibits more modest variations, not exceeding 0.2 eV. This should serve as an initial indication of the role played by the equatorial ligands in determining these quantities. Moreover, when comparing the two models utilizing a triple-zeta basis set, the observed trends remain consistent with those at the double-zeta level, showing average differences of less than 0.1 eV.
  
  %This result confirms the earlier observation of delocalization of high-lying orbitals in the bare uranyl ion when employing a smaller basis set.
  
  In comparison to the supermolecule model (computed at the double zeta level), the differences between the embedded uranyl and this model are relatively small, ranging from 1.3 to 3.9 eV. However, the differences between the bare uranyl and the supermolecular calculations can be as large as 24.4 eV. This finding underscores the capability of the FDE method to provide a reliable description of such a complex system in a computationally efficient manner.
  
  It is worth mentioning that we found that this strategy slightly overestimated the results for the valence-level excitations and ionizations in the supermolecule model by less than 2 eV, which exhibited good agreement with experimental observations \cite{gomes2013towards}. For XANES simulations, both models have demonstrated quantitative agreement in 4c-DR-TD simulations at the U M$_4$-edge. Moreover, both 4c-DR-TD and 4c-DR-TD-in-DFT  simulations exhibited similar deviations from the experiment (approximately 37 eV) \cite{misael2023core}, which can be attributed to the poor description of orbital correlation within DFT methods.
  
  Additionally, it is important to note that the binding energies of U $3d$ will not be further discussed in this work. As highlighted by \citeauthor{fujimori2019manifestation}\cite{fujimori2019manifestation}, these energies fall outside the range detectable by most X-ray photoelectron spectroscopy (XPS) instruments, which typically operate within the energy range of 20-1500 eV \cite{hufner2013photoelectron}. Therefore, the values presented in this study should be considered as a purely qualitative estimate for U $3d$ binding energies.
  
\begin{table*}%[ht]
  \caption{CVS-EOM-IP CBEs (in eV) for the models (a) \ce{UO2^{2+}}, (b) \ce{UO2^{2+}} @ \ce{Cl4^{4-}}, and (c) \ce{UO2Cl4^{2-}} are reported using double- and triple-zeta basis sets for all atoms. All calculations were performed with the $^4$DC Hamiltonian. When applicable, energy corrections due to QED effects and the Breit interaction reported by \citeauthor{koziol2018qed}~\cite{koziol2018qed} were added.}
\label{table*:cvs-eom-ipmodels}
%\resizebox{\textwidth}{!}{%
\begin{tabular}{llcccccc}
\hline\hline
\multicolumn{2}{c}{Assignment} &  QED+Breit & \multicolumn{5}{c}{ 4c-CVS-EOM-IP + QED + Breit} \\
& & \multicolumn{4}{c}{double-zeta} & \multicolumn{2}{c}{triple-zeta} \\
\cline{4-8}
& & & \ce{UO2^{2+}} & \ce{UO2^{2+}} @ \ce{Cl4^{4-}} & \ce{UO2Cl4^{2-}} & \ce{UO2^{2+}} @ \ce{Cl4^{4-}} & \ce{UO2^{2+}} \\
\hline
U & $3d_{3/2}$ & 8.34 & 3757.24 & 3736.97 & 3734.36 & 3736.80 & 3756.90 \\
&   $3d_{5/2}$ & 5.35 & 3579.57 & 3559.37 & 3556.76 & 3559.21 & 3579.31 \\
&    $4s$ & 5.37 & 1475.18 & 1455.24 & 1452.35 & 1455.13 & 1475.27 \\
&   $4p_{1/2}$ & 4.98 & 1308.26 & 1287.97 & 1285.11 & 1287.87 & 1308.02 \\
&  $4p_{3/2}$ & 2.72 & 1076.00 & 1055.75 & 1052.85 & 1055.63 & 1075.80 \\
&  $4d_{3/2}$ & 1.34 & 810.58 & 790.16 & 787.27 & 790.01 & 810.19 \\
&   $4d_{5/2}$ & 0.67 & 767.49 & 747.08 & 744.19 & 746.92 & 767.10 \\
&   $4f_{5/2}$ & - & 416.48 & 396.32 & 393.43 & 396.22 & 416.41 \\
&    $4f_{7/2}$ & - & 405.25 & 385.10 & 382.20 & 385.00 & 405.18 \\
\hline
O&    $1s$ & - & 555.52 & 534.29 & 533.00 & 534.24 & 555.41 \\
\hline\hline
\end{tabular}%}
\end{table*}
  
  \subsubsection{Comparison to experiment}
  \label{sec:exp}
  
%  Before proceeding, it is important to note that due to constraints in time and resources, the following study could not be extended to consider the entire crystalline environment of the \ce{Cs2UO2Cl4} system. Therefore, the discussion that follows is limited to the available simulations and their relative positions to experimental data. However, it is planned to further extend this study in order to provide a more comprehensive analysis of the system.

The influence of the crystalline environment on the chemical shifts associated with the binding energies of the uranyl ion becomes apparent upon comparison with experimental data. In Table \ref{table*:comp-exp}, we present our CVS-EOM-IP results with those obtained by XPS, as reported by \citeauthor{teterin2016valence}~\cite{teterin2016valence}, for various binding energies in the \ce{Cs2UO2Cl4} crystal.
   
  Our results, obtained using more elaborate models, demonstrate a trend towards better agreement with XPS data. For the bare uranyl ion, the calculated binding energies show relative positions ranging from 23 eV to 31 eV compared to the XPS data. In the case of the supermolecule calculation, all values, except for the U $4p_{3/2}$ binding energy which deviates by 8 eV, are higher than the XPS data by only 4.2 eV. As discussed in the previous section, the values obtained using the FDE scheme closely resemble those obtained for \ce{UO2Cl4^{2-}}, being higher than the XPS data by 6.5 eV for most of the cases, and 10.8 eV higher than the value reported for the U $4p_{3/2}$ binding energy.
  
  Here we would like to highlight several key points. Firstly, based on our benchmark of relativistic Hamiltonians, we anticipate that the calculated binding energies will show only modest variations when small-type integrals are explicitly computed. Specifically, at the double zeta level, these variations were found to be around 1.43 eV. Therefore, in addition to considering these integrals and making further advancements in electronic correlation and basis sets, it is expected, at the current stage of this work, that the rest of the crystalline environment -- e.g. cesium atoms in the vicinity of the uranyl ion -- to play a non-negligible role in determining the core-level binding energies here studied. 
  
 % Based on the available data and the comparison with XPS values, it is anticipated that the U $4p_{3/2}$ binding energy, as reported in this study, will undergo a decrease of approximately 11 eV when the remaining crystalline environment is included in the simulations. Similarly, we expect a relatively modest impact on the binding energies of U $4f_{5/2}, 4f_{7/2}$, and O $1s$, with an expected change of less than 3 eV compared to our current data. The subsequent work will adopt a similar approach to the one previously employed for analyzing low-lying binding energies \cite{gomes2013towards}. Specifically, two additional models will be considered: one involving the inclusion of the cesium atoms, and another incorporating the rest of the crystalline environment.
  
\begin{figure}%[!h]
    \centering
    \includegraphics[width=0.5\textwidth]{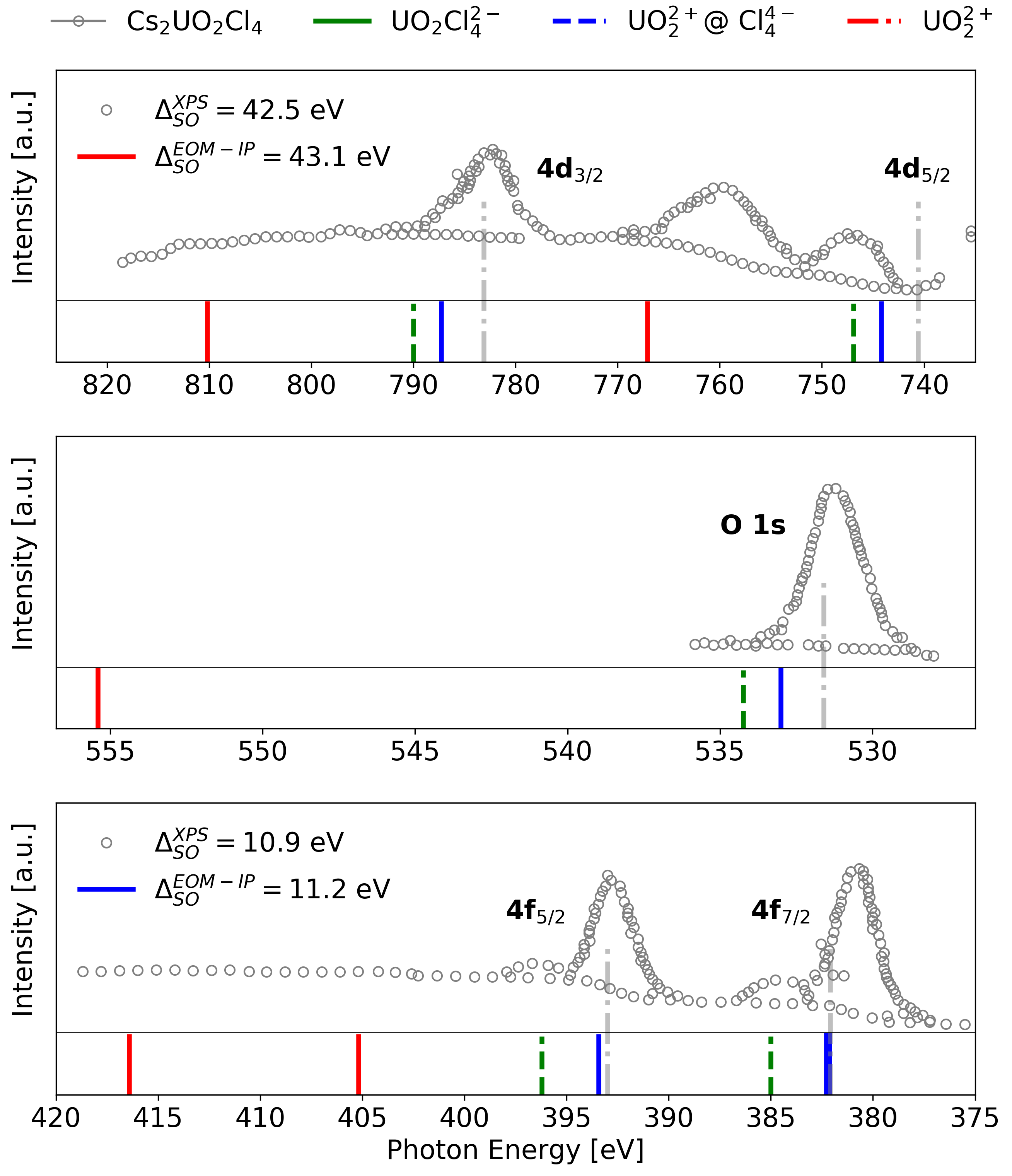}
    \caption{CVS-EOM-IP core ionization energies and spin-orbit splittings ($\Delta_{SO}$) in eV for \ce{UO2^{2+}} (red), \ce{UO2^{2+}} @ \ce{Cl4^{4-}} (blue), and \ce{UO2Cl4^{2-}} (green) using double- and triple-zeta basis sets for all atoms. These calculations were performed with the $^4$DC Hamiltonian. When applicable, energy corrections due to QED effects and the Breit interaction reported by \citeauthor{koziol2018qed}~\cite{koziol2018qed} were added. Experimental data from\citet{teterin2016valence}.}
    \label{fig:comp-teterin}
  \end{figure}

  In addition to that, our results provide further insights into the photoemission spectra reported by \citeauthor{teterin2016valence}, exhibiting considerable agreement with their estimation for the spin-orbit doublet in the U $4d$ XPS. We calculated the value as 43.09 eV, which differs by only 0.59 eV from their estimated value. This discrepancy is likely attributable to the presence of shake-up structures and the Cs $3d_{3/2}$ features, which complicate the experimental attribution and are not accounted for in our protocol as well. In the case of U $4f$ doublet, our calculations demonstrate an even smaller deviation in the spin-orbit doublet, amounting to only 0.33 eV. The success of our methodology is also evident in the determination of the $4p_{3/2}$ binding energy, which also exhibits a complex structure and could not be experimentally determined. Based on our results, we expect the spin-orbit doublet for U $4p$ XPS to be around 232.3 eV. The results of our CVS-EOM-IP calculations alongside the experimental values reported by \citeauthor{teterin2016valence} are depicted in Figure \ref{fig:comp-teterin}.
  
  \begin{table*}%[!h]
    \centering
    \caption{CVS-EOM-CCSD CBEs and spin-orbit splittings ($\Delta_{SO}$) in eV for models (a) \ce{UO2^{2+}}, (b) \ce{UO2^{2+}} @ \ce{Cl4^{4-}}, and (c) \ce{UO2Cl4^{2-}} using double- and triple-zeta basis sets for all atoms. The calculations were performed with the $^4$DC Hamiltonian. When applicable, energy corrections due to QED effects and the Breit interaction reported by \citeauthor{koziol2018qed}~\cite{koziol2018qed} were added. Experimental data from \citet{teterin2016valence}.}
    \label{table*:comp-exp}
    \begin{tabular}{lcccc}
  \hline\hline
      Assignment & UO$^{2+}_{2}$ & UO$^{2+}_{2}$ @ Cl$_{4}$ & UO$_{2}$Cl$_{4}$ & Cs$_2$UO$_{2}$Cl$_{4}$ (XPS) \\
  \hline
      U $4p_{1/2}$ & 1308.02 & 1287.87 & 1285.11 & - \\
      U $4p_{3/2}$ & 1075.80 & 1055.63 & 1052.85 & 1044.90 \\
      $\Delta_{SO}$ & 232.22 & 232.24 & 232.26 & - \\
      \hline
      U $4d_{3/2}$ & 810.19 & 790.01 & 787.27 & 783.10 \\
      U $4d_{5/2}$ & 767.10 & 746.92 & 744.19 & 740.60 \\
      $\Delta_{SO}$ & 43.09 & 43.09 & 43.08 & 42.50 \\
      \hline
      U $4f_{5/2}$ & 416.41 & 396.22 & 393.43 & 393.00 \\
      U $4f_{7/2}$ & 405.18 & 385.00 & 382.20 & 382.10 \\
      $\Delta_{SO}$ & 11.23 & 11.23 & 11.23 & 10.90 \\
      \hline
      O $1s$ & 555.41 & 534.24 & 533.00 & 531.60 \\
      \hline\hline
    \end{tabular}
  \end{table*}
  
  \section{Conclusions}
  \label{sec:conclusions}
  
  %\noindent
  
We have investigated, to the best of our knowledge for the first time in the literature with the relativistic CVS-EOM-CC approach, the core-level ionization energies of the uranyl ion in the \ce{Cs2UO2Cl4} crystal. Our study involved comparing different relativistic Hamiltonians, considering the role of dynamical correlation and orbital relaxation, and accounting for basis set truncation effects in the computation of core-level binding energies. Through these efforts, we have obtained valuable insights into the core-level spectroscopic observables of the uranyl ion.
  
  Building upon previous studies, we addressed the limitations associated with applying a 2-component Hamiltonian and adhering to the zeroth order truncation in electron-electron repulsion. Concerning the former, our simulations that employed a 2-component Hamiltonian showed small variations for those in which a 4-component framework was used. By including the Gaunt interaction in our calculations we could provide additional insights into the crucial role of the spin-other-orbit interaction on the reduction of the $2p$ binding energies.  
  
By improving the description of electronic correlation in our simulations, we showed the liability of applying the CVS-EOM-CC framework to investigate systems that include atoms at the bottom of the periodic table. In our results, the difference between CVS-EOM-IP and Koopmans' theorem didn't show significant changes when applying different Hamiltonians. The alignment of this result with the observed small differences between 2- and 4-component calculations supports the conclusion that 2-component calculations can be used without compromising the quantitative accuracy of the simulations.

We also compared different structural models, including the bare uranyl ion (\ce{UO2^{2+}}), the uranyl ion in the embedding potential of four chlorides (\ce{UO2^{2+}} @ \ce{Cl4^{4-}}), and the uranyl tetrachloride dianion (\ce{UO2Cl4^{2-}}). This comparison allowed us to analyze significant changes in binding energies attributed to the equatorial ligands near the uranyl ion. Furthermore, by comparing the embedded uranyl and supermolecular systems, we identified systematic differences with relatively small variations in binding energies, deviating by only 2.6 eV. By capturing non-negligible contributions across all binding energies investigated, these findings highlight the effectiveness of the FDE method in accurately recovering pivotal electrostatic interactions.
  
Concerning the experimental data, our methodology was particularly successful in determining the $4d$ and $4f$ spin-orbit splittings. By employing our protocol, we were able to yield results that closely align with those reported in the literature for the \ce{Cs2UO2Cl4} crystal, deviating by less than 0.7 eV from these values. This outcome is noteworthy, as the experimental determination of these doublets is challenging due to obscuration from partial cross-section contributions arising from secondary processes, such as satellite peaks and ligand excitations.

  In line with our simulations of the XANES of uranyl using the 4c-DR-TD-DFT-in-DFT method, these findings provide further support for our claim that the embedded uranyl serves as a cost-effective model for the uranyl tetrachloride system. Moreover, they highlight the crucial role of electrostatic interactions with the equatorial ligands in determining the spectroscopic observables of the uranyl unit.
  
  %It is anticipated that our results to be improved in a subsequent investigation when we finally proceed to include cesium atoms and the rest of the crystalline environment of \ce{Cs2UO2Cl4} in the calculations using the here employed FDE scheme. We expect that this will lower our actual binding energies up to \qty{10}{\electronvolt}.
  
  Given the broad range of applications of photoemission spectroscopies in studying actinides in condensed phases and the significant influence of their surrounding environment on determining chemical shifts, we foresee the need for further exploration of the current protocol in more complex systems. However, it is crucial to acknowledge that applying the CVS-EOM-CC method to larger systems, such as the \ce{Cs2UO2Cl4} crystal, remains elusive. In this context, we believe that the CVS-EOM-CC-in-DFT method used in this study offers a valuable approach for exploring the core-excited states of complex systems through reliable \textit{ab initio} molecular relativistic electronic structure calculations.

\section{Author contributions}

Conceptualization, methodology: W.A.M. and A.S.P.G; software: W.A.M and A.S.P.G; data curation, visualization: W.A.M; funding acquisition and resources: A.S.P.G; investigation, writing (original draft preparation): W.A.M; supervision: A.S.P.G; writing (review \& editing): W.A.M. and A.S.P.G.

\section{Acknowledgements}

W.A.M and A.S.P.G acknowledge support from the Franco-German project CompRIXS (Agence nationale de la recherche ANR-19-CE29-0019, Deutsche Forschungsgemeinschaft JA 2329/6-1), PIA ANR project CaPPA (ANR-11-LABX-0005-01), I-SITE ULNE projects OVERSEE and MESONM International Associated Laboratory (LAI) (ANR-16-IDEX-0004), the French Ministry of Higher Education and Research, region Hauts de France council and European Regional Development Fund (ERDF) project CPER CLIMIBIO, and the French national supercomputing facilities (grants DARI A0090801859, A0110801859, A0130801859). 

\clearpage

\bibliography{main.bib}

%\bibliographystyle{plainnat} % Use your preferred style
%\bibliography{main} % This links to references.bib

\end{document}